\documentclass[a4paper,twocolumn,fleqn]{article}

\usepackage{amsmath}             % recommended for advanced math
\usepackage{txfonts}               % free Times-based fonts for text and math
\usepackage{cite}                  % Ref. [1--3] instead of [1,2,3]
\usepackage[dvips]{graphicx}     % dvips
\usepackage{pfr_p}                   % PFR style
\usepackage{color}

\begin{document}
\newcommand{\aka}[1]{\textcolor{red}{#1}}
\title{Benchmark of the bootstrap current simulation in helical plasmas}

% Use \sup{1},... for superscripts in the authors and affiliations lines.

\author{Botsz HUANG\sup{1}, Shinsuke SATAKE\sup{1,2}, Ryutaro KANNO\sup{1,2}, Hideo SUGAMA\sup{1,2} \\ and Takuya GOTO\sup{1,2}}

\affiliation{
  \sup{1}Sokendai (The Graduate University for Advanced Studies), 509-5292, Toki, Gifu, Japan\\
  \sup{2}National Institute for Fusion Science, National Institutes of National Sciences, 509-5292, Toki, Gifu, Japan\\
  }

\date{(Received 5 December 2016)}
%\date{(Received 10 January 2009 / Accepted 24 January 2009)}

\email{huang.botsz@nifs.ac.jp}

\begin{abstract}
%The series of the benchmarks demonstrate the importance of the parallel momentum conservation on the bootstrap current evaluation in the non-axisymmetric sytem by the Zero-orbit-width(ZOW)\cite{Matsuoka2015}, the PENTA\cite{penta_code_2005} and the Drift-kinetic-equation solver (DKES)\cite{Rij_1989} models.
The importance of the parallel momentum balance %conservation% 
on the bootstrap current evaluation in non-axisymmetric systems is demonstrated by the benchmarks among the local drift-kinetic equation solvers, i.e., the Zero-Orbit-width (ZOW) model, DKES, and  PENTA.
The ZOW model is extended to include the ion parallel mean flow effect on the electron-ion parallel friction.
Compared to %the DKES model%
DKES code in which only the pitch-angle-scattering term is included in the collision operator, %the PENTA model %
PENTA  code employs the Sugama-Nishimura method to correct the momentum balance. 
The ZOW model and PENTA codes, both of which conserve the parallel momentum in like-species collisoins 
and include the electron-ion 
parallel frictions, agree each other well on the calculations of the bootstrap current. 
The DKES results without the parallel momentum conservation deviates significantly from those from the ZOW model
and PENTA. This work verifies the reliability of the bootstrap current calculation with the ZOW model and PENTA for the helical plasmas. 
\end{abstract}


%\DOI{10.1585/pfr.4.000}

\maketitle  % Don't forget to put this!

%%%%% MAIN TEXT %%%%%
The study of the bootstrap current is necessary to reproduce accurately the MHD equilibrium for high-beta plasmas.  
For the axisymmetric magnetic geometry, reliable analytic formulas of bootstrap current is available\cite{Sauter1999}.
For the non-axisymmetric system, one needs to rely on numerical methods to evaluate the bootstrap current, which is complicatedly dependent on the magnetic geometry, the collision frequency, and the radial electric field.
The past studies\cite{ISAEV2012} presented the benchmark between the Monte-Carlo global model VENUS+$\delta$f and the local semi-analytical solution SPBSC\cite{K_Watanabe_1992} in LHD.
The bootstrap current between the VENUS+$\delta$f and the SPBSC codes shows a systematic difference.
Although the difference may be caused in part by the finite-orbit-width effect, 
a missing discussion in that paper is about the treatment of collision term. The VENUS+$\delta$f code did not treat 
the friction force between electrons and ions, while SPBSC solved the balance between parallel viscosity and friction force 
as shown in later in Eq. \eqref{eq:parallel_momentum_balance} by analytic formula. 
%The large difference in the treatment of drift-kinetic equation in these two codes makes it difficult to study whether 
%the parallel friction is important or not for the bootstrap current simulation. 
In order to carry out a more direct investigation on the impact of the parallel friction on the bootstrap current calculations, this 
paper performs the benchmark among the ZOW model\cite{Matsuoka2015}, DKES\cite{Rij_1989}, and PENTA\cite{spong_2005},
which are all based on local neoclassical models.
%Furthermore, the ZOW model is carried out by the Monte-Carlo method and the $\delta$-f.
%The DKES model  is implemented by the Principle of Virtual Work.
%Then, the numerical bootstrap current is discussed as follows.  

The ZOW model\cite{huang_2016} solves the radially-local drift-kinetic equation by the $\delta f$ Monte-Carlo method, and the parallel friction $\mathcal { F }_{\parallel}$ is treated as follows.
For the like-species collisions, the linearized collision operators are employed and this satisfies the parallel momentum balance, i.e., $\mathcal { F }_{\parallel, ee} =\mathcal { F }_{\parallel, ii} = 0$.
For ion, the ion-electron friction $\mathcal{ F }_{\parallel,ie}$ is neglected because of the large mass ratio, $m_{e}/ m_{i} \ll 1$.
For electron, in the previous work, the electron-ion collision was only approximated as the pitch-angle scattering operator with the stationary background Maxwellian ion distribution,  i.e., $\mathcal{ C }_{ei}\simeq \mathcal{ L }_{ei}$.
In the present work, not only the pitch-angle scattering but also the ion parallel mean flow $U_{\parallel,i}$ are newly employed,
\begin{equation} \label{eq:fei_mean_flow}
 \mathcal{ C }_{ei} \cong  \mathcal{ L }_{ei} + \nu_{D}^{ei} \frac{ m_e }{ T_e} U_{\parallel,i} \nu_{\parallel} f_{e \mathcal{M} }.
\end{equation}
With the new $\mathcal{ C }_{ei}$ operator, the electrons are exposed to the friction $\mathcal{ F}_{\parallel, ei}$ which is roughly proportional to $(  U_{\parallel,i} - U_{\parallel,e})$.
In Eq.\eqref{eq:fei_mean_flow}, the ion parallel mean flow $U_{\parallel,i}$ is given as
\begin{equation}\label{eq:v_parallel_i}
 U_{\parallel,i} 
 = \frac{ \left\langle U_{\parallel,i} B \right\rangle }{ \left\langle B^2 \right\rangle } B
 + \left( \frac{1}{en}  \frac{ dp_i (\psi) }{ d\psi } + \frac{ d\Phi (\psi) }{ d\psi } \right) \widetilde{ U}_{\parallel}.
\end{equation}
where $\langle \cdots \rangle$ represents a flux-surface average, and the pressure $ p_i (\psi) $ and the 
electrostatic potential $\Phi (\psi)$ are assumed as the flux-surface functions. The second term in Eq. \eqref{eq:v_parallel_i} 
represents the return flow of the diamagnetic and $E\times B$ flow, with the assumption that these flows are 
divergence-free on the flux-surface\cite{spong_2005}.
The $\widetilde{ U}_{\parallel}$ term vanishes after taking the flux-surface average, i.e., 
$\left \langle B \widetilde{ U}_{\parallel} \right \rangle = 0.$
%Thus, this term is neglected in Eq.\eqref{eq:fei_mean_flow}.
In Eq.\eqref{eq:v_parallel_i}, the term  $\left\langle U_{\parallel,i} B \right\rangle $ is given from the ion simulations. 

DKES solves the local and mono-energy drift-kinetic equation.
Both ions and electrons implement the pitch-angle scattering in the their collision operators
$
 \mathcal{C}_a \cong \sum_{b} \mathcal{L}_{a,b}. 
$
Therefore, the momentum balance is not accurately satisfied either in the like- or the unlike-species collision.
%Besides the original DKES in the work, the DKES-like model is employed which is the DKES model with the same collision operators as the ZOW model. 
In PENTA code\cite{spong_2005}, Sugama-Nishimura method\cite{penta_theory_2002} is adapted in order to re-interpret the diffusion coefficients from DKES so that the momentum conservation is satisfied,
i.e., $\mathcal { F }_{\parallel, ii} = \mathcal { F }_{\parallel,ee}=0 $ and $\mathcal { F }_{\parallel, ei} = -\mathcal { F }_{\parallel, ie}$.
%so that the conservation law is satisfied in the parallel momentum balance equation, i.e., $\mathcal { F }_{\parallel, ii} = -\mathcal { F }_{\parallel, ee} = 0 $ and $\mathcal { F }_{\parallel, ei} = -\mathcal { F }_{\parallel, ie}$.
%Ideally, the PENTA model reproduces the intrinsic-ambipolarity in an axisymmetric system.
The exact momentum balance as in the Sugama-Nishimura method is essential to reproduce 
the intrinsic ambipolarity in the axisymmetric limit\cite{penta_theory_2002}. 
Besides the collision operator, the main difference in the ZOW model and 
DKES/PENTA is the guiding-center motion in the local approximation. While both the $E\times B$ and the magnetic drift 
terms tangential to the flux-surface are retained in the ZOW model, the magnetic drift is neglected and
the incompressible-$E\times B$ approximation is used in DKES and PENTA\cite{huang_2016}.
 
\begin{figure}
   \includegraphics[width=0.45\textwidth]{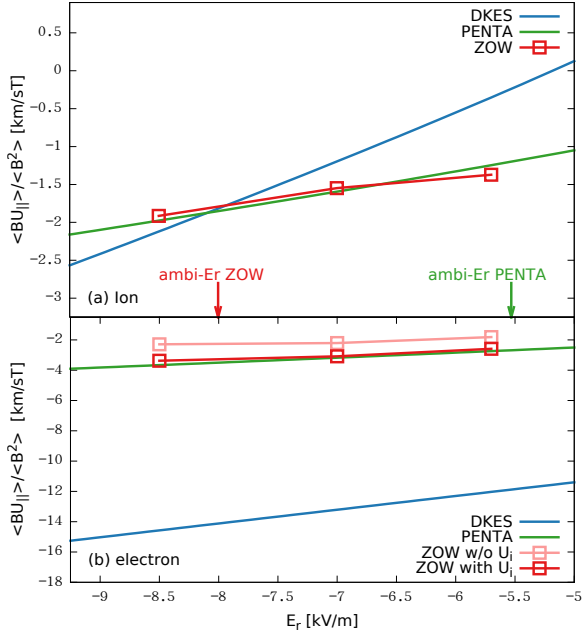}
 \caption{The dependence of the flux-surface average parallel mean flow  $\langle BU_{\parallel}\rangle/\langle B^2\rangle$
 on radial electric field : (a) deuterium ion and (b) electron. 
 The ``ambi-$E_r$'' labels represent the vaules of ambipolar-$E_r$ evaluated from the ZOW and PENTA simulations.}
  \label{fig:fei_U}
\end{figure}

In Fig.\ref{fig:fei_U}, the parallel flows from the ZOW model, DKES, and PENTA are presented under the condition considered as a self-ignition operation point of FFHR-d1\cite{goto_2015}.
The parallel mean flow is determined by the following momentum balance equation,
\begin{align} \label{eq:parallel_momentum_balance}
 \left \langle \frac{ \partial }{ \partial t } mn U_{\parallel} B \right \rangle
   + \left \langle \boldsymbol{B} \cdot \nabla \cdot \left( \boldsymbol{P}_{\text {CGL} } + \boldsymbol{\Pi}_{2 } \right) \right \rangle
 = \left \langle\mathcal { F }_{\parallel} B \right \rangle.
\end{align}
Here $\boldsymbol{P}_{\text {CGL} }$ is the diagonal viscosity tenser.
$\boldsymbol{\Pi}_{2 }$ is the non-diagonal viscosity tensor which is related to the parallel  and 
$\boldsymbol E \times \boldsymbol B$ flows\cite{huang_2016}.
Note that the $\widetilde{ U}_{\parallel,i}$ term in Eq.\eqref{eq:v_parallel_i} has no contribution to 
$\langle\mathcal { F }_{\parallel} B \rangle$ because $\left \langle B \widetilde{ U}_{\parallel} \right \rangle = 0.$
Following Eq.\eqref{eq:parallel_momentum_balance} and the assumption of small impact of the friction on the ion momentum balance, the steady-state ion paralell mean flow is determined so that the total ion parallel viscosity vanishes;
\begin{equation}\label{eq:parallel_momentum_balance_ion}
 \left \langle \boldsymbol{B} \cdot \nabla \cdot \left( \boldsymbol{P}_{\text {CGL} } + \boldsymbol{\Pi}_{2 } \right) \right \rangle _{i} \simeq 0.
\end{equation}
In Fig.\ref{fig:fei_U}(a), the ZOW model and PENTA agree with each other well even though the $\mathcal{F}_{\parallel,ie}$ is absent in the ZOW model.
This suggests that the friction $\mathcal{F}_{\parallel,ie}$ is in fact negligible as it is expected 
from the large mass ratio $m_e/m_i\ll 1$.
%This suggests that the friction $\mathcal{F}_{\parallel,ie}$ is negligible as expected in the ZOW model in the Sec.\ref{sec:friction_intro}.
The gap between the results from DKES and PENTA indicates that it is necessary to maintain the momentum conservation 
of the like-species collisions even in the helical plasmas. 
For the electrons, in Fig.\ref{fig:fei_U}(b), there are two results from the ZOW model in order to examine the impact of 
the ion parallel mean flow $U_{\parallel,i}$.
The electron parallel momentum equation depends on the balance,
\begin{equation}\label{eq:parallel_momentum_balance_electron}
 \left \langle \boldsymbol{B} \cdot \nabla \cdot \left( \boldsymbol{P}_{\text {CGL} } + \boldsymbol{\Pi}_{2 } \right) \right \rangle_{e} \simeq \left \langle\mathcal { F }_{\parallel,ei} B \right \rangle.
\end{equation}
%In Fig.\ref{fig:fei_U}(b),  in the friction $\mathcal{F}_{\parallel,ei}$.
In Fig.\ref{fig:fei_U}(b), the friction $\mathcal{F}_{\parallel,ei}$  with the finite $U_{\parallel,i}$ gives rise to the gap between the two ZOW simulation results.
The ZOW result with finite $U_{\parallel,i}$ agrees with the PENTA's one. 
These models both maintain the parallel momentum conservation with finite $U_{\parallel,i}$ and the momentum correction of the like-species collision, respectively.
For the ZOW model, it is obvious that the correct ion parallel mean flow is necessary to improve 
the collision operator on the electron parallel flow calculation. 
For the electron, there is also the large gap between the results from DKES and PENTA as in the ion simulations.

\begin{figure}
   \includegraphics[width=0.45\textwidth]{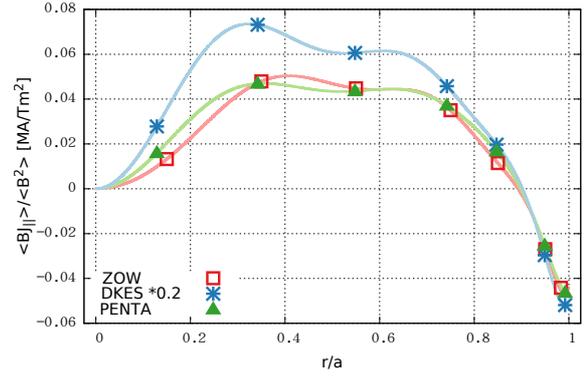}
 \caption{The radial profile of the bootstrap current in the FFHR-d1 case at the ambipolar condition. 
 The DKES result is multiplied by 0.2. The ZOW calculation 
 includes the finite-$U_{\parallel.i}$ effect in the friction force $\mathcal { F }_{\parallel, ei}$.}
 \label{fig:vb_ambi}
\end{figure}

In Fig.\ref{fig:vb_ambi}, the radial profile of the bootstrap current in the FFHR-d1 case  at the ambipolar condition 
is estimated by the three codes. 
The bootstrap current from the ZOW model with finite $U_{\parallel,i}$ agrees with PENTA. In the previous studies\cite{Matsuoka2015, huang_2016} it is found that neglecting the 
tangential magnetic drift in DKES and PENTA causes the overestimation of the ion radial particles flux when $E_r$ is small. 
This results in the difference in the ambipolar-$E_r$ values as shown in Fig.\ref{fig:fei_U}.
However, in the present case, since $\langle BJ_\parallel\rangle =\langle Bne(U_{\parallel,i}-U_{\parallel,e})\rangle$ from 
the ZOW model and PENTA have very weak dependence on $E_r$, the bootstrap current from these two codes agrees each other.
%At $r/a > 0.95$, the difference between the ZOW and PENTA is attributed to the boundary of the curve-fitting.
%This boundary is given as the same as on the last flux surface, because the ZOW model does not evaluate the flux surface on the edge.
The DKES result shows approximately 10 times larger magnitude of the bootstrap current than those from PENTA and the ZOW models.

It is well-known that the pitch-angle scattering operator is enough to evaluate the radial neoclassical fluxes in helical plasmas.
However, it is insufficient  for the bootstrap current calculation.
The present study shows that both the momentum conservation in the like-species collision and the friction acting on the electrons are important physics to estimate the bootstrap current correctly, even in helical plasmas. 
%The collision operator is implemented as Eq.\eqref{eq:fei_mean_flow} in the ZOW model.
%Besides the pitch-angle scattering, the proper friction is necessary to improve the collision operator of the simulation, especially for the parallel electron mean flow.
%This work shows the verification of the difference bootstrap current calculations and it indicates that the parallel momentum balance is important in not only the tokamak but also the stellarator. 
This work is also the first report of verification of the ZOW model and PENTA for bootstrap current calculations. 
These two codes will serve to improve the accuracy of the bootstrap current calculation in general helical plasmas.
\bibliographystyle{unsrt}
%\bibliography{paper2.bib}

\end{document}